\documentclass[11pt,a4paper]{article}
\usepackage{amsmath,amsfonts,amssymb,amscd,graphicx}
\catcode`\@=11
\@addtoreset{equation}{section}

\catcode`\@=12

\newtheorem{Proposition}{Proposition}[section]

\newtheorem{Remark}{Remark}[section]

\newfont{\gotico}{eufm10 scaled\magstephalf}
\newfont{\qvd}{msam10 scaled\magstephalf}

\def\demo{\par\noindent{\sc Proof. }\begingroup}
\def\enddemo{\hskip1em \mbox{\qvd \char3}\endgroup\par\medskip}

\def\interior{\,\hbox{\vrule depth0pt height.6pt width4pt%
\vrule depth0pt height8pt}\;\,}

\def\de#1/de#2{\frac{\partial {#1}}{\partial {#2}}}
\def\De#1/de#2{\dfrac{\partial {#1}}{\partial {#2}}}

\def\det{{\rm det}\,}

\def\const{{\rm const.}}

\def\jE{{\cal J}\/({\cal E})}
\def\E{{\cal E}}

\begin{document}
\vskip-2cm

\title{A vielbein formulation of unified Einstein--Maxwell theory}

\author{Stefano Vignolo\\
        DIPTEM Sez. Metodi e Modelli Matematici, Universit\`a di Genova \\
                Piazzale Kennedy, Pad. D - 16129 Genova (Italia)\\ 
                E-mail: vignolo@diptem.unige.it
\and
       Enrico Massa \\
        Dipartimento di Matematica, Universit\`a di Genova \\
        Via Dodecaneso 35 - 16146 Genova (Italia) \\
                E-mail: massa@dima.unige.it
}

\date{}              
\maketitle

\begin{abstract}{\noindent In the framework of $\cal J$-bundles a vielbein formulation of unified Einstein--Maxwell theory is proposed.
In the resulting scheme, field equations matching the gravitational and electromagnetic fields are derived by constraining a $5$-dimensional variational principle. No dynamical scalar field in involved.}
\end{abstract}

\section{Introduction}
Most of the existing field theories admit a variational formulation, developed on a suitable first jet--bundle. However, in many cases --- for example in Gauge Theory and in General Relativity --- the corresponding Lagrangian density is singular. Often, this is due to the fact that the Lagrangian depends on the partial field derivatives only through suitable antisymmetric combinations (e.g. the Lagrangian of the electromagnetic field).

This is indicative of the fact that the fibre coordinates of the entire jet bundle are redundant for these theories. In other words, they represent too many degrees of freedom, without any direct physical interpretation. 

These considerations have been the starting point for the definition of the $\cal J$-bundles and the study of their geometry \cite{CVB1,CVB2,VC,VCB2,VCB1}. The idea is to consider a suitable quotient space of the first jet--bundle, making two sections equivalent when they possess a first order contact with respect to the exterior (or exterior covariant) differentiation, rather than with respect to the whole set of derivatives. The fibre coordinates of the resulting quotient space (the $\cal J$-bundle) are just the antisymmetric combinations of the field derivatives appearing in the Lagrangian. 

The so--defined $\cal J$-bundles have been used to set up a new formulation of Gauge theories and General Relativity in the Poincar\'e--Cartan formalism. The relevant choices for the fibre $\cal J$-coordinates have been shown to be the components of the strength tensor for Yang--Mills theories \cite{CVB1,CVB2,VCB2} and the torsion and curvature tensors for General Relativity \cite{VC}. Such approach has resulted in cutting away some unphysical degrees of freedom (represented by unnecessary jet--coordinates) and in reducing the degeneracy of Yang--Mills theories \cite{CVB1,CVB2,VCB2} and General Relativity \cite{VCB1}.

In this paper a further aspect of the relationship between $\cal J$-bundles and field theories is discussed, namely the interaction between the geometrical construction proposed in \cite{VCB1} and Kaluza's unified description of Einstein--Maxwell theory.

This aim is achieved by first extending the purely--frame formulation of General Relativity given in \cite{VCB1} to a $5$-dimensional principal bundle $Q\/$ over the space-time and then constraining the variational principle yielding Einstein equations in vacuo.

The resulting scheme allows to represent interacting gravitational and electromagnetic fields as a pseudo--riemannian metric on $Q\/$ described in terms of vielbeins. More in detail, the electromagnetic potential $A=A_i\,dx^i\/$ and the tetrad elements $e^\mu = e^\mu_i\,dx^i\/$ are joined together and are seen to give rise to the orthonormal vielbeins for the metric on $Q\/$. An analogous result, derived from the Jordan--Thiry theory in the purely metric formulation, can be found in \cite{L}.   

The Einstein and Maxwell theories present a great structural similarity: both may be described in terms of $1$-forms defined on the space--time manifold; in both cases, the field equations are of second--order for the dynamical fields ($A\/$ and $e\/$). Taking all this into account, the proposed geometrical approach allows to unify the above theories in a simple and self--contained mathematical setting.

The paper is organized as follows. In Section 2 we extend the geometrical construction proposed in \cite{VCB1} to arbitrary $m$-dimensional manifolds.

In Section 3 we apply the mathematical machinery to a suitable $5$-dimensional principal fibre bundle $Q\/$, deriving Einstein--Maxwell equations from a constrained variational principle.

In this respect, we notice that in the previous paper \cite{VCB1} we worked within the gauge natural bundle framework \cite{FF}. The latter provides the suitable mathematical setting for globally describing gravity in the tetrad formalism, without any topological restriction on the nature of space--time (such as parallelizability). Global aspects are clearly important when conservation laws and first integrals are considered, due to the intrinsically non--local nature of these objects.

In the present paper, we shall not discuss these topics, but shall focus attention on the differential equations of the theory and on their derivation. Therefore, for simplicity, we have chosen to work in the more standard natural bundle framework. As a consequence, up to the parallelizability hypothesis, the variational principle we used has a local nature.

\section{$\cal J\/$-bundles and General Relativity}
Let $M\/$ be an orientable manifold of dimension $m$, allowing a metric tensor $g\/$ of signature $\eta=(p,q)=(\underbrace{-1,\ldots,-1}_{p\;\; times},\underbrace{1,\ldots,1}_{q\;\; times})\/$. We set $\eta_{\mu\nu}=\eta^{\mu\nu}:=diag\;(\underbrace{-1,\ldots,-1}_{p\;\; times}\/$, $\underbrace{1,\ldots,1}_{q\;\; times})\/$ $\mu,\nu =1,\ldots,m\/$.

Let $\E :=L^*\/(M)\/$ be the co--frame bundle over $M\/$, referred to local coordinates $x^i,e^\mu_i\/$ $i,\mu =1,\ldots,m\/$. Local sections $e:M\to\E\/$ are identified with local co--frames on $M\/$ expressed as $e^\mu\/(x) = e^\mu_i\/(x)\,dx^i\/$.

Since our aim is developing a suitable geometrical description of General Relativity in terms of co--frame fields, the metric of $M\/$ will be described by means of a family of local sections (the local orthonormal co--frames) $e:M\to\E\/$, defined modulo the action of the group $SO\/(p,q)\/$ and glued to each other by Lorentz transformations.

By construction the theory has then to be invariant under two groups of transformations, namely Lorentz transformations and coordinate transformations. The action of both these groups on $\E\/$ is locally expressed as 
\begin{equation}\label{2.1}
\bar{x}^j = \bar{x}^j\/(x^i),\qquad \bar{e}^\mu_j = e^\sigma_i
\Lambda^\mu_{\;\;\sigma}\/(x) \de x^i/de{\bar{x}^j}
\end{equation}
with $\Lambda^\mu_{\;\;\sigma}\/(x)\in SO\/(p,q)\;\forall\;x\in M$. Transformations (\ref{2.1}) will henceforth be referred to as {\it gauge transformations\/}. 

Let us now focus on the first jet bundle $j_1\/(\E)\/$ associated with the fibration $\pi:\E\to M\/$. We refer $j_1\/(\E)\/$ to local jet--coordinates $x^i,e^\mu_i,e^\mu_{ij}\/$. 

The geometrical construction proposed in \cite{VCB1} relies on the introduction of a suitable equivalence relation on $j_1\/(\E)\/$: two elements $z\/$ and $\hat z\/$ of $j_1\/(\E)\/$, projecting onto the same $x\in M\/$, are said equivalent if and only if  
\begin{equation}\label{2.2}
e^\mu\/(x) =  \hat{e}^\mu\/(x) \quad {\rm and } \quad de^\mu\/(x) =  d\hat{e}^\mu\/(x)
\end{equation}
$e^\mu\/$ and $\hat{e}^\mu\/$ denoting any two sections of $\pi:\E\to M\/$ chosen among the representatives of the equivalence classes $z\/$ and $\hat z\/$ respectively. In other words, two sections $e^\mu\/$ and $\hat{e}^\mu\/$ are regarded as equivalent when they possess a first order contact with respect to the exterior differentiation, rather than with respect to the whole set of derivatives. The above equivalence relation is geometrically well defined, since it is easily recognized to be independent of the choice of the representatives $e^\mu\/$ and $\hat{e}^\mu\/$ belonging to the classes $z\/$ and $\hat z\/$.

In local coordinates, if $z=(x^i,e^\mu_i,e^\mu_{ij})\/$ and $\hat{z}=(x^i,\hat{e}^\mu_i,\hat{e}^\mu_{ij})\/$, it is immediately seen that $z\sim \hat{z}\/$ if and only if the following relations holds
\begin{equation}\label{2.3}
    e^\mu_i =  \hat e^\mu_i \quad {\rm and } \quad (e^\mu_{ij} - e^\mu_{ji}) =
    (\hat e^\mu_{ij} - \hat e^\mu_{ji})
\end{equation}
We denote by $\jE:=j_1\/(E)/\sim\/$ the quotient space and by $\rho:j_1\/(E)\to\jE\/$ the corresponding quotient map. A system of local fibered coordinates on the bundle $\jE$ is provided by
$x^i,e^\mu_i,E^\mu_{\;ij}:=\frac{1}{2}\left(e^\mu_{ij} - e^\mu_{ji} \right) (i<j)$.

The geometry of $\cal J$-bundles has been thoroughly examined in Refs. \cite{CVB1,CVB2,VC,VCB2}. 
As proved there, the quotient map $\rho\/$ endows $\jE$ with most of the standard features of jet--bundles geometry: $\cal J$-extensions of sections, contact forms, $\cal J$-prolongations of morphisms and vector fields.

In particular, gauge transformations (\ref{2.1}) may be $\cal J$-prolongated to the bundle $\jE\/$; their $\cal J$-prolongations are described locally by eqs.~(\ref{2.1}) together with (see \cite{VCB1} and references therein for details) 
\begin{equation}\label{2.4}
\bar{E}^\mu_{jk} = E^\sigma_{ih}\Lambda^\mu_{\;\;\sigma}\de x^h/de{\bar{x}^k}\de
x^i/de{\bar{x}^j} + \frac{1}{2}e^\sigma_i\de\Lambda^\mu_{\;\;\sigma}/de{x^h}\de
x^h/de{\bar{x}^k}\de x^i/de{\bar{x}^j} -
\frac{1}{2}e^\sigma_i\de\Lambda^\mu_{\;\;\sigma}/de{x^h}\de x^h/de{\bar{x}^j}\de
x^i/de{\bar{x}^k}
\end{equation}
where $E^\mu_{ij}:=-E^\mu_{ji}\/$ whenever $i>j\/$.

A suitable set of new coordinates may be now introduced on $\jE\/$. In fact, the components of the spin--connections generated by the co--frames themselves may be chosen as fibre coordinates on the bundle $\jE$. 

To see this point, let $z=(x^i,e^\mu_i,E^\mu_{\;ij})$
be an element of $\jE$, $x=\hat\pi\/(z)$ its projection over $M$ and $e^\mu$ a
representative co--frame belonging to the equivalence class $z$. The (local) co--frame $e^\mu\/$ defines a corresponding (local) metric $g=\eta_{\mu\nu}\,e^\mu \otimes e^\nu\/$ on $M$ which in turn induces a Levi--Civita connection $\Gamma_{ih}^k\/$. The latter, expressed in terms of the non--holonomic basis $e^\mu\/$, yields the coefficients of spin--connection $\omega_{i\;\;\;\nu}^{\;\;\mu}\/$.

The relation between the coefficients $\Gamma_{ih}^k$ of the Levi--Civita
connection and the coefficients $\omega_{i\;\;\;\nu}^{\;\;\mu}$ of the associated spin--connection,
evaluated at the point $x=\hat \pi\/(z)\in M$, is expressed by the equation
\begin{equation}\label{2.5}
\omega^{\;\;\mu}_{i\;\;\;\nu}\/(x) = e^\mu_k\/(x)\left( \Gamma^k_{ij}e^j_\nu\/(x) +
\de{e^k_\nu\/(x)}/de{x^i} \right)
\end{equation}

More specifically, if the coefficients
$\Gamma_{ih}^k$ are written in terms of the co--frame $e^\mu$ and its derivatives, one gets
the well--known expression
\begin{equation}\label{2.6}
\omega^{\;\;\mu}_{i\;\;\nu}\/(x)= e^\mu_p\/(x) \left( \Sigma^p_{\;\;ji}\/(x) -
\Sigma_{j\;\;i}^{\;\;p}\/(x) + \Sigma_{ij}^{\;\;\;p}\/(x) \right) e^j_{\nu}\/(x)
\end{equation}
with
\begin{equation}\label{2.7}
\Sigma^p_{\;\;ji}\/(x):= e^p_\lambda\/(x) E^\lambda_{ij}\/(x) =
e^p_\lambda\/(x)\frac{1}{2}\left( \de{e^\lambda_i\/(x)}/de{x^j} -
\de{e^\lambda_j\/(x)}/de{x^i} \right)
\end{equation}
the Latin indices being lowered and raised by means of the metric $g=\eta_{\mu\nu}\,e^\mu \otimes e^\nu\/$.

Equations \eqref{2.6} and \eqref{2.7} show that the values of the coefficients
of the spin--connection $\omega_{i\;\;\;\nu}^{\;\;\mu}\/$, evaluated at $x=\hat \pi\/(z)\/$,
are independent of the choice of the representative $e^\mu\/$ in the equivalence class
$z\in \jE\/$.

Moreover, the torsion--free condition for the connection
$\omega_{i\;\;\;\nu}^{\;\;\mu}$ gives a sort of inverse relation of eq.~(\ref{2.6}) in the form
\begin{equation}\label{2.8}
2E^\mu_{ij}\/(x) = \de{e^\mu_i\/(x)}/de{x^j} -
\de{e^\mu_j\/(x)}/de{x^i} = \omega^{\;\;\mu}_{i\;\;\;\nu}\/(x)e^\nu_j\/(x) -
\omega^{\;\;\mu}_{j\;\;\;\nu}\/(x)e^\nu_i\/(x)
\end{equation}

Because of the metric compatibility condition $\omega_i^{\;\;\mu\nu}:=
\omega_{i\;\;\;\sigma}^{\;\;\mu} \eta^{\sigma\nu} = - \omega_i^{\;\;\nu\mu}$, there exists
a one-to-one correspondence between the values of the antisymmetric part of the
derivatives $E^\mu_{ij}\/(x) = \frac{1}{2} \left( \de e^\mu_{i}\/(x) /de{x^j} - \de e^\mu_{j}\/(x)
/de{x^i}\right)$ and the coefficients of the spin--connection $\omega_i^{\;\;\mu\nu}\/(x)$ at
the point $x=\hat\pi\/(z)$.

The above considerations allow us to take the quantities
$\omega_i^{\;\;\mu\nu}$ ($\mu < \nu\/$) as fibre coordinates of the bundle $\jE$, looking at the
relations \eqref{2.6} and \eqref{2.8} as coordinate changes in $\jE$.

From eqs. (\ref{2.4}) and (\ref{2.8}), or also directly from eq. (\ref{2.5}), it is easily seen that in the coordinates $x,e,\omega\/$ the ($\cal J$-prolongations of) gauge transformations on $\jE\/$ are described by eq. (\ref{2.1}) together with 
\begin{equation}\label{2.9}
\bar{\omega}^{\;\;\mu\nu}_{i}=\Lambda^\mu_{\;\;\sigma}\/(x)\Lambda^{\nu}_{\;\;\gamma}\/(x)\de
x^j/de{\bar{x}^i}\omega^{\;\;\sigma\gamma}_{j} - \Lambda_{\sigma}^{\;\;\eta}\/(x)\de
\Lambda^\mu_{\;\;\eta}\/(x)/de{x^h}\de x^h/de{\bar{x}^i}\eta^{\sigma\nu}
\end{equation}
where $\Lambda_{\sigma}^{\;\;\nu}:=\Lambda^{\alpha}_{\;\;\beta}\eta_{\alpha\sigma}\eta^{\beta\nu}=\left(\Lambda^{-1}\right)^\nu_{\;\;\sigma}\/$ and $\omega_i^{\;\;\mu\nu}:=-\omega_i^{\;\;\nu\mu}\/$ whenever $\mu>\nu\/$. Through eqs. \eqref{2.9} we recover the well known transformation laws of the spin--connection coefficients. 

For convenience of the reader, we recall also the concepts of contact forms and $\cal J\/$-extensions of sections. 

Given a section $e:M\to\E\/$, we define its $\cal J\/$-extension as ${\cal J}\/e := \rho\circ j_1\/e\/$, namely as the projection of the jet--extension $j_1\/e\/$ on $\jE\/$ through the quotient projection $\rho\/$.

A section $\gamma :M\to\jE\/$ is said holonomic if it is the $\cal J\/$-extension of some section $e:M\to\E\/$, i.e. $\gamma = {\cal J}\/e\/$. In local coordinates, a section $\gamma\/$ is holonomic if and only if $\gamma : x \mapsto \left( x^i,e^\mu_i\/(x),E^\mu_{ij}\/(x) = \frac{1}{2} \left( \de e^\mu_{i}\/(x) /de{x^j} - \de
e^\mu_{j}\/(x) /de{x^i}\right)\right)$.

As it happens for standard jet--bundles, the bundle $\jE\/$ is endowed with suitable contact forms. The latter are $2$-forms locally spanned by the $m$ $2$-forms
\begin{equation}\label{2.9bis}
\theta^\mu := de^\mu_i \wedge dx^i + E^\mu_{ij}\,dx^i\wedge dx^j = de^\mu_i \wedge dx^i + \omega_{i\;\;\;\nu}^{\;\;\mu}e^\nu_j\,dx^i\wedge dx^j
\end{equation}
Under gauge transformations, the $2$-forms \eqref{2.9bis} transform as 
\begin{equation}
\bar{\theta}^\mu = \Lambda^\mu_{\;\;\sigma}\theta^\sigma
\end{equation}
It is immediately seen that a section $\gamma :M\to\jE\/$ is holonomic if and only if $\gamma^*\/(\theta^\mu)=0\/$ $\forall\; \mu =1,\ldots,m\/$. 

Our plan is now to derive the field equations for General Relativity from a variational principle stated on the manifold $\jE\/$.

To this end, we introduce an $m$-form on $\jE\/$, locally described as
\begin{equation}\label{2.10}
\Theta := \frac{1}{(N-2)!2}\epsilon^{qp\cdot\cdot\cdot lij}\epsilon_{\mu\nu\cdot\cdot\cdot\rho\lambda\sigma}\/e^\mu_{q}e^\nu_{p}\cdot\cdot\cdot e^\rho_l\left( d\omega_{i}^{\;\;\lambda\sigma} \wedge ds_j + \omega_{j\;\;\;\eta}^{\;\;\lambda}\omega_{i}^{\;\;\eta\sigma}\,ds \right)
\end{equation}
where $ds:=dx^1 \wedge\ldots\wedge dx^m\/$, $ds_i := \de /de{x^i}\interior ds\/$ and $\epsilon\/$ denotes the Levi--Civita permutation symbol. (In $m=4\/$ dimensions and in a quite different geometrical setting, the same form \eqref{2.10} has been used in \cite{Rovelli}). The following result holds true
\begin{Proposition}\label{Pro2.1}
The form $\Theta\/$ (\ref{2.10}) is invariant under gauge transformations
(\ref{2.1}), (\ref{2.9}) on the manifold ${\cal J}\/(\cal E)\/$.
\end{Proposition}
\demo
As for $m=4\/$ dimensions \cite{VCB1}, the proof consists in a direct check, taking eqs. \eqref{2.1}, \eqref{2.9} and the identities
\[
\begin{split}
\bar{\omega}_{j\;\;\;\eta}^{\;\;\tau}\bar{\omega}_{i\;\;\;\sigma}^{\;\;\eta} = \Lambda^{\tau}_{\;\;\alpha}\Lambda_{\sigma}^{\;\;\beta}\de{x^k}/de{\bar{x}^j}\de{x^h}/de{\bar{x}^i}\omega_{k\;\;\;\lambda}^{\;\;\alpha}\omega_{h\;\;\;\beta}^{\;\;\lambda} +
\Lambda^{\tau}_{\;\;\alpha}\de{\Lambda_{\sigma}^{\;\;\beta}}/de{x^h}\de{x^h}/de{\bar{x}^i}\de{x^k}/de{\bar{x}^j}\omega_{k\;\;\;\beta}^{\;\;\alpha} +\\
-\de{\Lambda_{\;\;\alpha}^{\tau}}/de{x^h}\de{x^h}/de{\bar{x}^j}\Lambda_{\sigma}^{\;\;\beta}\de{x^k}/de{\bar{x}^i}\omega_{k\;\;\;\beta}^{\;\;\alpha} -
\de{\Lambda_{\;\;\eta}^{\tau}}/de{x^h}\de{x^h}/de{\bar{x}^j}\de{\Lambda_{\sigma}^{\;\;\eta}}/de{x^k}\de{x^k}/de{\bar{x}^i}
\end{split}
\]
into account.
\enddemo 
By means of the $m$-form $\Theta\/$ we may define the (local) action functional
\begin{equation}\label{2.11}
A\/(\gamma):= \int_D \gamma^*\/(\Theta)
\end{equation}
for every section $\gamma : D\subset M\to {\cal J}\/(\cal E)\/$, $D\/$ compact domain.

The study of the stationarity conditions for the functional (\ref{2.11}) leads to the Einstein equations in the vielbein formulation. To see this point we recall that, by imposing usual vanishing conditions at the boundary $\partial D\/$, a section $\gamma: x\to (x^i,e^\mu_i\/(x),\omega_i^{\;\;\mu\nu}\/(x))\/$ is critical for $A\/$ if and only if it satisfies the condition (see, for example, \cite{Hermann})
\begin{equation}\label{2.12}
\gamma^*\/(X\interior d\Theta)=0
\end{equation}
for every vector field $X= X^\mu_i\,\de /de{e^\mu_i} + \frac{1}{2}X_i^{\mu\nu}\,\de
/de{\omega_{i}^{\;\;\mu\nu}}\/$ on ${\cal J}\/(\cal E)\/$ (with $X_i^{\mu\nu}=-X_i^{\nu\mu}\/$ when $\mu > \nu\/$) vertical with respect to the fibration $\jE\to M\/$. Now
\begin{equation}\label{2.13}
\begin{split}
d\Theta = \frac{1}{(N-3)!2}\epsilon^{qp\cdot\cdot\cdot lij} \epsilon_{\mu\nu\cdot\cdot\cdot\rho\lambda\sigma}e^\mu_{q}e^\nu_{p}\cdot\cdot\cdot\,de^\rho_{l}\wedge\left( d\omega_{i}^{\;\;\lambda\sigma} \wedge ds_j + \omega_{j\;\;\;\eta}^{\;\;\lambda}\omega_{i}^{\;\;\eta\sigma}\,ds \right) +\\ 
+\frac{1}{(N-2)!}\epsilon^{qp\cdot\cdot\cdot lij} \epsilon_{\mu\nu\cdot\cdot\cdot\rho\lambda\sigma}\/e^\mu_{q}e^\nu_{p}\cdot\cdot\cdot e^\rho_{l}\omega_{j\;\;\;\eta}^{\;\;\lambda}\,d\omega_{i}^{\;\;\eta\sigma} \wedge ds
\end{split}
\end{equation}
Moreover, generalizing \cite{VCB1} to $m$ dimensions, we have 
\begin{Proposition}\label{Pro2.2}
the following identity
\begin{equation}\label{2.14}
\begin{split}
\frac{1}{(N-2)!}\epsilon^{qp\cdot\cdot\cdot lij}\epsilon_{\mu\nu\cdot\cdot\cdot\rho\lambda\sigma}\/e^\mu_{q}e^\nu_{p}\cdot\cdot\cdot e^\rho_{l}\omega_{j\;\;\;\eta}^{\;\;\lambda}\,d\omega_{i}^{\;\;\eta\sigma} =\\
-\frac{1}{(N-3)!2}\epsilon^{qp\cdot\cdot\cdot lij}\epsilon_{\mu\nu\cdot\cdot\cdot\tau\lambda\sigma}\/e^\mu_{q}e^\nu_{p}\cdot\cdot\cdot e^\rho_{l}\omega_{j\;\;\;\rho}^{\;\;\tau}\,d\omega_{i}^{\;\;\lambda\sigma}
\end{split}
\end{equation}
holds. 
\end{Proposition}
\demo
The proof is easily obtained by direct computation, inserting the identities $\omega_{j\;\;\;\eta}^{\;\;\lambda}\,d\omega_{i}^{\;\;\eta\sigma}=\omega_{s\;\;\;\eta}^{\;\;\lambda}\,d\omega_{t}^{\;\;\eta\sigma}e^s_{\beta}e^\beta_{j}e^t_{\alpha}e^\alpha_i\/$ and $\omega_{j\;\;\;\rho}^{\;\;\tau}\,d\omega_{i}^{\;\;\lambda\sigma}=\omega_{s\;\;\;\rho}^{\;\;\tau}\,d\omega_{t}^{\;\;\lambda\sigma}e^s_{\beta}e^\beta_{j}e^t_{\alpha}e^\alpha_i\/$ in eq. \eqref{2.14}
\enddemo
Taking the identity (\ref{2.14}) into account, we may rewrite the differential of $\Theta\/$ in the form 
\begin{equation}\label{2.15}
\begin{split}
d\Theta = \frac{1}{(N-3)!2}\epsilon^{qp\cdot\cdot\cdot lij} \epsilon_{\mu\nu\cdot\cdot\cdot\rho\lambda\sigma}e^\mu_{q}e^\nu_{p}\cdot\cdot\cdot\,de^\rho_{l}\wedge\left( d\omega_{i}^{\;\;\lambda\sigma} \wedge ds_j + \omega_{j\;\;\;\eta}^{\;\;\lambda}\omega_{i}^{\;\;\eta\sigma}\,ds \right)+\\ 
-\frac{1}{(N-3)!2}\epsilon^{qp\cdot\cdot\cdot lij}\epsilon_{\mu\nu\cdot\cdot\cdot\rho\lambda\sigma}\/e^\mu_{q}e^\nu_{p}\cdot\cdot\cdot e^\tau_{l}\omega_{j\;\;\;\tau}^{\;\;\rho}\,d\omega_{i}^{\lambda\sigma} \wedge ds
\end{split}
\end{equation}
Given a vertical vector field $X= X^\mu_i\,\de /de{e^\mu_i} + \frac{1}{2}X_i^{\mu\nu}\,\de
/de{\omega_{i}^{\;\;\mu\nu}}\/$ on ${\cal J}\/(\cal E)\/$, one has then
\begin{equation}\label{2.16}
\begin{split}
X\interior d\Theta = \frac{1}{(N-3)!2}\epsilon^{qp\cdot\cdot\cdot lij}\epsilon_{\mu\nu\cdot\cdot\cdot\rho\lambda\sigma}\/e^\mu_{q}e^\nu_{p}\cdot\cdot\cdot\,\left( d\omega_{i}^{\;\;\lambda\sigma} \wedge ds_j + \omega_{j\;\;\;\eta}^{\;\;\lambda}\omega_{i}^{\;\;\eta\sigma}\,ds \right)\/X^\rho_{l} +\\
- \frac{1}{(N-3)!2}\epsilon^{q\cdot\cdot\cdot lij}\epsilon_{\mu\nu\cdot\cdot\cdot\rho\lambda\sigma}\/e^\mu_{q}e^\nu_{p}\cdot\cdot\cdot\,\left( de^\rho_l \wedge ds_j + e^\tau_{l}\omega_{j\;\;\;\tau}^{\;\;\rho}\,ds \right)\/X^{\lambda\sigma}_{i}
\end{split}
\end{equation}
In view of this and due to the arbitrariness of the vector fields $X\/$, the condition (\ref{2.12}) gives rise to two sets of final equations, respectively expressed as
\begin{subequations}
\begin{equation}\label{2.17a}
\frac{1}{(N-3)!}\epsilon^{qp\cdot\cdot\cdot lij}\epsilon_{\mu\nu\cdot\cdot\cdot\rho\lambda\sigma}\/e^\mu_{q}e^\nu_{p}\cdot\cdot\cdot\/\left( \de{e^\rho_l}/de{x^j} + \omega_{j\;\;\;\tau}^{\;\;\rho}e^\tau_{l} \right) = 0
\end{equation}
and
\begin{equation}\label{2.17b}
\frac{1}{(N-3)!2}\epsilon^{qp\cdot\cdot\cdot lij}\epsilon_{\mu\nu\cdot\cdot\cdot\rho\lambda\sigma}\/e^\mu_{q}e^\nu_{p}\cdot\cdot\cdot\/\left( \de{\omega_{i}^{\;\;\lambda\sigma}}/de{x^j} + \omega_{j\;\;\;\eta}^{\;\;\lambda}\omega_{i}^{\;\;\eta\sigma} \right) = 0
\end{equation}
\end{subequations}
At this point, it is a straightforward matter to verify that eqs.~(\ref{2.17a}) are equivalent to the relations
\begin{equation}\label{2.18}
2E^\nu_{pj}\/(x) = \omega_{p\;\;\;\rho}^{\;\;\nu}\/(x)e^\rho_{j}\/(x) - \omega_{j\;\;\;\rho}^{\;\;\nu}\/(x)e^\rho_{p}\/(x) = \de{e^\nu_p}/de{x^j}\/(x) - \de{e^\nu_j}/de{x^p}\/(x)
\end{equation}
which ensure the kinematic admissibility of critical sections or, in other words, the fact that the quantities $\omega_{i}^{\;\;\mu\nu}\/(x)\/$ are exactly the coefficients of the spin connection associated with the Levi--Civita connection induced by the metric $g_{ij}\/(x)=\eta_{\mu\nu}e^\mu_{i}\/(x)e^\nu_j\/(x)\/$ on $M\/$. As a result, eqs.~(\ref{2.17b}) are then identical to Einstein equations in $m$ dimensions (of course, provided that $\det\/(e^\mu_i)\not = 0\/$)
\begin{equation}\label{2.19}
\frac{1}{(N-3)!4}\epsilon^{qp\cdot\cdot\cdot lij}\epsilon_{\mu\nu\cdot\cdot\cdot\rho\lambda\sigma}\/e^\mu_{q}\/(x)e^\nu_{p}\/(x)\cdot\cdot\cdot\/R_{ji}^{\;\;\;\lambda\sigma}\/(x)=0
\end{equation}
$R_{ji}^{\;\;\;\lambda\sigma}\/(x):= \de{\omega_{i}^{\;\;\lambda\sigma}}/de{x^j}\/(x) - \de{\omega_{j}^{\;\;\lambda\sigma}}/de{x^i}\/(x) +
\omega_{j\;\;\;\eta}^{\;\;\lambda}\/(x)\omega_{i}^{\;\;\eta\sigma}\/(x) - \omega_{i\;\;\;\eta}^{\;\;\lambda}\/(x)\omega_{j}^{\;\;\eta\sigma}\/(x)\/$ denoting the curvature tensor of the metric $g\/$.

It is worth noticing that, in the case of matter coupling, the formalism can allow for connections with non--vanishing torsion, so describing an Einstein--Cartan like theory. Of course, in such a circumstance, critical sections $\gamma: M\to \jE\/$ are no longer holonomic; to restore holonomy, one should suitably modify the transformation laws \eqref{2.6} and \eqref{2.8}.

\section{Affine scalar bundle and Einstein--Maxwell theory}
In this Section we show that the geometrical construction illustrated in Section 2 allows a unified formulation of the Einstein--Maxwell theory. 

To start with, let $M\/$ be a $4$-dimensional orientable space--time manifold, allowing a metric tensor of signature $(1,3)=(-1,1,1,1)\/$. We denote by $\eta_{\mu\nu}=\eta^{\mu\nu}=diag\,(-1,1,1,1)\/$ ($\mu,\nu =1,\ldots 4)\/$.

Let us consider a principal fibre bundle $\pi : Q\to M\/$ over $M\/$, with structural group $(\Re,+)\/$. We shall call $Q\/$ the bundle of affine scalars over $M\/$. Fibre coordinates on $Q\/$ are (local) functions $x^i,x^5\/$ ($i=1,\ldots,4\/$) where $x^i\/$ are coordinates on $M\/$ and $x^5\/$ is a trivialization of $\pi : Q\to M\/$.

In fibre coordinates, the vector field $\de /de{x^5}\/$ identifies with the generator of the $1$-parameter group of translations along the fibres of $Q\/$, usually referred to as the fundamental vector field of $Q\/$.

Still in fibre coordinates, equivariant diffeomorphisms on $Q\/$ are locally expressed as
\begin{equation}\label{4.1}
    \left\{
\begin{array}{l}
\bar{x}^i = \bar{x}^i\/(x^j) \\
\\
\bar{x}^5 = x^5 + f\/(x^j)
\end{array}
\right.
\end{equation}
where $f\/(x^i) \in {\cal F}\/(M)\/$. Their Jacobians are described by 
\begin{equation}\label{4.2}
J = \de(\bar{x}^i,\bar{x}^5)/de{(x^j,x^5)}=
\left(
\begin{array}{ll}
\de{\bar{x}^i}/de{x^j} & 0 \\
\\
\de f/de{x^j} & 1
\end{array} 
\right)
\end{equation}
with $\det\/(\de{\bar{x}^i}/de{x^j})\not = 0\/$.
    
The first step in our plan consists in developing the formulation of General Relativity explained in the previous Sections, taking the principal fibre bundle $Q\/$ as the base manifold of the theory.

To this end, we use once again the co--frame bundle $\E :=L^*\/(Q)\/$ of $Q\/$ as the configuration space of our theory. We refer $\E\/$ to local coordinates $x^i,x^5,e^\mu_i,e^5_i,\/$ $e^\mu_5,e^5_5\/$ ($i,\mu=1,\ldots,4\/$). 

In this context, we attempt to achieve a geometric unified description of the interacting gravitational and electromagnetic fields by looking for a suitable family of pseudo--Riemannian geometries on the bundle $Q\/$.

More precisely, we want to single out those gravitational fields $g\/$ on $Q\/$ whose description in terms of (local) orthonormal co--frames is given by ``space--time'' tetrads $e^\mu\/(x^j) = e^\mu_i\/(x^j)\,dx^i\/$ on $M\/$, Lie--transported along the fundamental vector field, completed by a fifth element $e^5\/(x^j) = dx^5 -kA_i\/(x^j)\,dx^i\/$ ($k=\const\/$), representing geometrically a principal connection on the bundle $Q\to M\/$. 
Every such required metric tensor $g\/$ is then locally espressed as 
\begin{equation}\label{4.2bis}
g := \eta_{\mu\nu}\,e^\mu\otimes e^\nu + e^5\otimes e^5
\end{equation}
We notice that, by definition \eqref{4.2bis}, the horizontal distribution generated by the connection $e^5\/$ is everywhere orthogonal to the vertical bundle and that the fundamental vector field $\de /de{x^5}\/$ is a Killing vector field, whose integral curves form a rigid congruence of geodesics.

Every metric tensor (\ref{4.2bis}) induces an associated Lorentzian metric $\tilde{g}\/$ on the space--time manifold $M\/$, defined by $\pi^*(\tilde{g}) := g - e^5\otimes e^5\/$.

The $1$-form $A := e^5 -dx^5 = -kA_i\/(x^j)\,dx^i\/$ is indentified with an electromagnetic potential on $M\/$, scaled by a multiplicative factor $k\/$ (depending on the used units) in order to get the right coupling constant $k^2\/$ in the field equations. In this way, the vortex tensor $\Omega := -2de^5\/$ \cite{Massa} coincides, up to the multiplicative factor $-2k\/$, with the electromagnetic tensor $F=-\frac{1}{2}\/\left(\de A_i/de{x^j} - \de A_j/de{x^i}\right)\,dx^j\wedge dx^i\/$ associated with $A\/$.

By construction, all fields $e^\mu\/$ and $e^5\/$ satisfy the condition $L_{\de /de{x^5}}e^\mu = L_{\de /de{x^5}}e^5 =0\/$ ($\Leftrightarrow L_{\de /de{x^5}}\/(g)=0\/$). In the present case, this fact implies that the fields $e^\mu\/$ and $e^5\/$ have to be independent of the fifth variable $x^5\/$. This is nothing but a restatement in more geometrical terms of the well known ''Kaluza's cylinder condition'' (see, for example, \cite{OW}), amounting to the requirement that the dynamical fileds are sections of the fibration $\E\to M\/$ rather than $\E\to Q\/$.

As we shall see, such a geometrical construction results to be invariant under a sub-group of gauge transformations on $\E\/$, locally described by eqs.~(\ref{4.1}) completed with 
\begin{equation}\label{4.3}
\bar{e} = \Lambda\cdot e \cdot J^{-1}
\end{equation}
in which the $SO(1,4)$-matrices $\Lambda\/$ are of the form 
\begin{equation}\label{4.4}
\Lambda\/(x)=
\left(
\begin{array}{ll}
\Lambda^\mu_{\;\;\nu}\/(x) & 0 \\
\\
0 & 1
\end{array}
\right)
\end{equation}
with $\Lambda^\mu_{\;\;\nu}\/(x) \in SO\/(1,3) \quad\forall x\in M\/$. Taking eqs. (\ref{4.2}) and (\ref{4.4}) into account, eq. (\ref{4.3}) takes the explicit form
\begin{subequations}\label{4.5}
\begin{equation}
\bar{e}^\mu_j = \Lambda^\mu_{\;\;\sigma}e^\sigma_{i}\de{x^i}/de{\bar{x}^j} - \Lambda^\mu_{\;\;\sigma}e^\sigma_{5}\de f/de{\bar{x}^j}
\end{equation}
\begin{equation}
\bar{e}^5_j = e^5_{i}\de{x^i}/de{\bar{x}^j} - e^5_{5}\de f/de{\bar{x}^j}
\end{equation}
\begin{equation}
\bar{e}^5_5 = e^5_5
\end{equation}
\begin{equation}
\bar{e}^\mu_5 = \Lambda^\mu_{\;\;\sigma}e^\sigma_5
\end{equation}
\end{subequations}
The latter, together with eqs. (\ref{4.1}), represent the gauge transformations allowed by the theory.
 
The $\cal J$-bundle $\jE\/$ is now taken into account. Consistently with eqs. \eqref{2.6}, \eqref{2.7} we refer $\jE\/$ to local coordinates of the kind $x^i,x^5,e^\mu_i,e^5_i,e^\mu_5,e^5_5,\omega_i^{\;\;\mu\nu},\omega_i^{\;\;\mu5}\/$, $\omega_5^{\;\;\mu\nu},\omega_5^{\;\;\mu5}\/$. $\cal J$-prolongations on $\jE\/$ of gauge transformations (\ref{4.1}), (\ref{4.5}) are then expressed as (compare eq. (\ref{2.9}) with eqs.  (\ref{4.2}) and (\ref{4.4}))
\begin{subequations}\label{4.6}
\begin{equation}
\bar{\omega}_i^{\;\;\mu\nu} = \Lambda^\mu_{\;\;\sigma}\Lambda^\nu_{\;\;\gamma}\de{x^j}/de{\bar{x}^i}\omega_j^{\;\;\sigma\gamma} - \Lambda_\sigma^{\;\;\gamma}\de{\Lambda^\mu_{\;\;\gamma}}/de{x^h}\de{x^h}/de{\bar{x}^i}\eta^{\sigma\nu} - \Lambda^\mu_{\;\;\sigma}\Lambda^\nu_{\;\;\gamma}\de f/de{\bar{x}^i}\omega_5^{\;\;\sigma\gamma}
\end{equation}
\begin{equation}
\bar{\omega}_i^{\;\;\mu5} = \Lambda^\mu_{\;\;\nu}\de{x^j}/de{\bar{x}^i}\omega_j^{\;\;\nu5} - \Lambda^\mu_{\;\;\nu}\de f/de{\bar{x}^i}\omega_5^{\;\;\nu5}
\end{equation}
\begin{equation}
\bar{\omega}_5^{\;\;\mu5} = \Lambda^\mu_{\;\;\nu}\omega_5^{\;\;\mu5}
\end{equation}
\begin{equation}
\bar{\omega}_5^{\;\;\mu\nu} = \Lambda^\mu_{\;\;\sigma}\Lambda^\nu_{\;\;\gamma}\omega_5^{\;\;\sigma\gamma}
\end{equation} 
\end{subequations}
Taking the restrictions about the geometries of $Q\/$ into account, the field equations of the theory may now be derived through the next two steps:

\bigskip
\noindent
i) we constrain the variational principle, built through the form (\ref{2.10}) (specialized for $m=5\/$ dimensions), to the submanifold ${\cal A}\subset \jE\/$ expressed locally as
\begin{equation}\label{4.7}
e^5_5 =1 \qquad,\qquad e^\mu_5 =0
\end{equation}
ii) we impose that the reduced dynamical fields $e^\mu_i\/$ and $-kA_i := e^5_i\/$ be sections of the fibration $\E\to M\/$, i.e. that they do not depend on the variable $x^5\/$.

\bigskip
\noindent
In other words, first we define a variational principle on the submanifold $i: {\cal A}\to\jE\/$ through the pull--back $i^*\/(\Theta)\/$ of the form \eqref{2.10}; after that, we look for solutions $\gamma\/$ of the (reduced) associated Euler--Lagrange equations such that they are independend of the variable $x^5\/$, namely $\gamma :M\to {\cal A}\/$.

\bigskip
As we shall see below, conditions i) and ii) yield the Einstein--Maxwell equations for the space--time metric $\tilde{g}\/$ and the electromagnetic potential $A\/$.

\begin{Remark}\label{Rem2.1}
The constraint \eqref{4.7} has a holonomic nature, i.e. it does not involve any partial derivative of the fields. In this respect, we remark that we are not forced to consider the $\cal J$-prolongation of \eqref{4.7} in $\jE\/$ and work on it. Indeed, because of the regularity of the variational principle \eqref{2.12}, the critical sections are automatically $\cal J$-extensions, whose images belong to the $\cal J$-extension of \eqref{4.7}.  
Also, we notice that (compare with eqs. (\ref{4.5}c) and (\ref{4.5}d)) eqs. (\ref{4.7}) are invariant under gauge transformations, while eqs. (\ref{4.5}a) and (\ref{4.5}b) reproduce respectively Lorentz transformations for the tetrad $e^\mu\/$ and usual gauge transformations for the potential $A\/$. 
\end{Remark}
The Euler--Lagrange equations generated by the constrained variational principle are still of the form (\ref{2.12}), where now $\gamma: M\to {\cal A}\/$ and the infinitesimal deformations $X\/$ (corresponding to variations on $\cal A\/$) are forced to be tangent to the submanifold $\cal A\/$ itself, i.e. have the local expression
\begin{equation}\label{4.8}
X = X^\mu_i\,\de /de{e^\mu_i} + X_i\,\de /de{e^5_i} + X^\omega\,\de /de{\omega}
\end{equation}
$X^\omega\,\de /de{\omega}\/$ denoting synthetically the vertical part (with respect to $\jE\to\E\/$). As mentioned in Remark \ref{Rem2.1}, the latter remains totally free and its arbitrariness gives rise to the first set of final equations (\ref{2.17a}), ensuring the holonomy of the critical sections. According to ii), by inserting the explicit expression of a possible solution ($e^\mu_i\/(x^j),\/$ $e^5_i\/(x^j)=-kA_i\/(x^j), e^5_5\/(x^j)=1, e^\mu_5\/(x^j)=0\/$) into eqs. (\ref{2.18}) (equivalent to \eqref{2.17a}), we obtain then the complete characterization of the ``vertical part'' of the solution itself given by   
\begin{subequations}\label{4.9}
\begin{equation}
\omega_5^{\;\;\rho5}\/(x^j)=0
\end{equation} 
\begin{equation}
\omega_5^{\;\;\rho\lambda}\/(x^j)=-\frac{1}{2}kF^{\rho\lambda}\/(x^j)
\end{equation}
\begin{equation}
\omega_{j}^{\;\;\nu5}\/(x^j)=-\frac{1}{2}kF^\nu_{\;\;\rho}\/(x^j)e^\rho_j\/(x^j)
\end{equation}
\begin{equation}
\omega_i^{\;\;\mu\nu}\/(x^j)=\tilde{\omega}_{i}^{\;\;\mu\nu}\/(x^j) + \frac{1}{2}k^{2}F^{\mu\nu}\/(x^j)A_i\/(x^j)
\end{equation}
\end{subequations}
where $F_{\mu\nu}\/(x^k):=F_{ji}\/(x^k)e^j_{\mu}\/(x^k)e^i_{\nu}\/(x^k)\/$, $F_{ji}\/(x^k)=\de A_j/de{x^i}\/(x^k) - \de A_i/de{x^j}\/(x^k)\/$ is the electromagnetic tensor generated by $A\/$ and $\tilde{\omega}_{i}^{\;\;\mu\nu}\/(x^k)\/$ are the coefficients of the spin connection over $M\/$ induced by the space--time metric $\tilde{g}\/(x^k) = \eta_{\mu\nu}\,e^\mu\/(x^k)\otimes e^\nu\/(x^k)\/$. It is a straightforward matter to verify that the transformation laws of the quantities \eqref{4.9} are consistent with eqs.~\eqref{4.6}.

Eqs. (\ref{2.12}) and (\ref{4.8}) yield twenty further equations, given by (compare eq. (\ref{2.15}) with eq. (\ref{4.8}))    
\begin{subequations}\label{4.10}
\begin{equation}
\frac{1}{4}\epsilon^{qplij}\epsilon_{\mu\nu\rho\lambda\sigma}\/e^\mu_{q}e^\nu_{p}\/\left( \de{\omega_{i}^{\;\;\lambda\sigma}}/de{x^j} + \omega_{j\;\;\;\eta}^{\;\;\lambda}\omega_{i}^{\;\;\eta\sigma} \right) = 0\qquad\quad l\not=5,\rho\not=5
\end{equation}
\begin{equation}
\frac{1}{4}\epsilon^{qplij}\epsilon_{\mu\nu5\lambda\sigma}\/e^\mu_{q}e^\nu_{p}\/\left( \de{\omega_{i}^{\;\;\lambda\sigma}}/de{x^j} + \omega_{j\;\;\;\eta}^{\;\;\lambda}\omega_{i}^{\;\;\eta\sigma} \right) = 0\qquad\quad l\not=5
\end{equation}
\end{subequations}
For simplicity, in eqs. (\ref{4.10}) --- and only in these --- all Latin and Greek indices run from $1\/$ to $5\/$.

Inserting the result (\ref{4.9}) into eqs. (\ref{4.10}) and taking eqs. \eqref{4.7} as well as point ii) into account, after some direct calculations (see appendix A), eqs. (\ref{4.10}a) are recognized to be exactly the Einstein--Maxwell equations (all indices run once again from $1$ to $4$)  
\begin{equation}\label{4.11}
\frac{1}{4}\epsilon^{5plij}\epsilon_{5\nu\rho\lambda\sigma}e^\nu_{p}\tilde{R}_{ji}^{\;\;\;\lambda\sigma}=-\frac{1}{2}ek^2T^l_\rho 
\end{equation}
where $e:=\det\left(e^\mu_i\right)\/$ ($\mu,i =1,\ldots,4\/$), $T^l_\rho :=\frac{1}{4}e^l_{\rho}F_{ij}F^{ij} + F^l_{\;\;j}F^j_{\;\;i}e^i_{\rho}\/$ is the energy--momentum tensor of the electromagnetic field $F\/$ and  
$\tilde{R}_{ji}^{\;\;\;\lambda\sigma}:= \de{\tilde{\omega}_{i}^{\;\;\lambda\sigma}}/de{x^j} - \de{\tilde{\omega}_{j}^{\;\;\lambda\sigma}}/de{x^i} +
\tilde{\omega}_{j\;\;\;\eta}^{\;\;\lambda}\tilde{\omega}_{i}^{\;\;\eta\sigma} - \tilde{\omega}_{i\;\;\;\eta}^{\;\;\lambda}\tilde{\omega}_{j}^{\;\;\eta\sigma}\/$ denotes the curvature tensor of the space--time metric $\tilde{g}=\eta_{\mu\nu}\,e^\mu \otimes e^\nu\/$.

With analogous calculations (appendix A), eqs. (\ref{4.10}b) are rewritten as
\begin{equation}\label{4.12}
\frac{1}{2}eke^i_{\beta}\/\left(\de{F^{\alpha\beta}}/de{x^i} + \tilde{\omega}_{i\;\;\;\eta}^{\;\;\alpha}F^{\eta\beta} + \tilde{\omega}_{i\;\;\;\eta}^{\;\;\beta}F^{\alpha\eta}\/\right)=0 
\end{equation}
identical, up to the inessential factor $\frac{1}{2}ek\/$, to the dynamical Maxwell equations
\begin{equation}\label{4.13}
\nabla_{\beta}F^{\alpha\beta}=0
\end{equation}
Eqs. (\ref{4.11}) and (\ref{4.12}) express the coupling of the gravitational field $\tilde{g}\/(x^k)=\eta_{\mu\nu}\,e^\mu\/(x^k)\otimes e^\nu\/(x^k)\/$, with the electromagnetic field induced by the potential $A\/(x^k)\/$.

\appendix
\section{Appendix}
For convenience of the reader, we show the basic steps involved in the derivation of equations \eqref{4.11} and \eqref{4.12}. We start by rewriting eqs. (\ref{4.10}a) 
\begin{equation}\label{A.1}
\frac{1}{4}\epsilon^{qplij}\epsilon_{\mu\nu\rho\lambda\sigma}\/e^\mu_{q}e^\nu_{p}\/\left( \de{\omega_{i}^{\;\;\lambda\sigma}}/de{x^j} + \omega_{j\;\;\;\eta}^{\;\;\lambda}\omega_{i}^{\;\;\eta\sigma} \right) = 0\qquad\quad l\not=5,\rho\not=5
\end{equation}
where all indices run from $1$ to $5$. Taking the constraint \eqref{4.7} as well as the point ii) into account, after some algebraic simplifications we get the expressions
\begin{equation}\label{A.2}
\begin{split}
\frac{1}{2}\epsilon^{5plij}\epsilon_{5\nu\rho\lambda\sigma}\/\left( \de{\omega_{i}^{\;\;\lambda\sigma}}/de{x^j} + \omega_{j\;\;\;\eta}^{\;\;\lambda}\omega_{i}^{\;\;\eta\sigma} \right)\/e^\nu_{p} + 
\frac{1}{2}\epsilon^{5plij}\epsilon_{5\nu\rho\lambda\sigma}\omega_{j\;\;\;5}^{\;\;\lambda}\omega_{i}^{\;\;5\sigma}e^\nu_{p} + \\
\frac{1}{2}\epsilon^{5plij}\epsilon_{5\nu\rho\lambda\sigma}\/\left( \de{\omega_{5}^{\;\;\lambda\sigma}}/de{x^j} + \omega_{j\;\;\;\eta}^{\;\;\lambda}\omega_{5}^{\;\;\eta\sigma} \right)\/e^5_{p}e^\nu_i 
- \frac{1}{2}\epsilon^{5plij}\epsilon_{5\nu\rho\lambda\sigma}\omega_{5\;\;\;\eta}^{\;\;\lambda}\omega_{j}^{\;\;\eta\sigma}e^5_{p}e^\nu_i + \\
+ \frac{1}{2}\epsilon^{5plij}\epsilon_{5\nu\rho\lambda\sigma}\omega_{5\;\;\;\eta}^{\;\;\sigma}\omega_{j}^{\;\;\eta5}e^\nu_{p}e^\lambda_i =0
\end{split}
\end{equation}
in which Latin and Greek indices run now from $1$ to $4$. By inserting the result \eqref{4.9} in \eqref{A.2}, we obtain
\begin{equation}\label{A.3}
\begin{split}
\frac{1}{2}\epsilon^{5plij}\epsilon_{5\nu\rho\lambda\sigma}\/\left[\de /de{x^j}\/\left(\tilde{\omega}_{i}^{\;\;\lambda\sigma} + \frac{1}{2}k^2F^{\lambda\sigma}A_i\right) + \left(\tilde{\omega}_{j\;\;\;\eta}^{\;\;\lambda} + \frac{1}{2}k^2F^\lambda_{\;\;\eta}A_j\right)\/\left(\tilde{\omega}_{i}^{\;\;\eta\sigma} + \frac{1}{2}k^2F^{\eta\sigma}A_i\right)\right]\/e^\nu_p + \\
+ \frac{1}{2}\epsilon^{5plij}\epsilon_{5\nu\rho\lambda\sigma}\/\left(-\frac{1}{2}kF^\lambda_{\;\;\alpha}e^\alpha_j\right)\/\left(\frac{1}{2}kF^\sigma_{\;\;\beta}e^\beta_j\right)\/e^\nu_p+\\
+
\frac{1}{2}\epsilon^{5plij}\epsilon_{5\nu\rho\lambda\sigma}\/\left[\de /de{x^j}\/\left(-\frac{1}{2}kF^{\lambda\sigma}\right) + \left(\tilde{\omega}_{j\;\;\;\eta}^{\;\;\lambda} + \frac{1}{2}k^2F^\lambda_{\;\;\eta}A_j\right)\/\left(-\frac{1}{2}kF^{\eta\sigma}\right)\right]\/\left(-kA_p\right)e^\nu_i +\\
- \frac{1}{2}\epsilon^{5plij}\epsilon_{5\nu\rho\lambda\sigma}\/\left(-\frac{1}{2}kF^\lambda_{\;\;\eta}\right)\/\left(\tilde{\omega}_{j}^{\;\;\eta\sigma} + \frac{1}{2}k^2F^{\eta\sigma}A_j\right)\/\left(-kA_p\right)e^\nu_i + \\
+
\frac{1}{2}\epsilon^{5plij}\epsilon_{5\nu\rho\lambda\sigma}\/\left(-\frac{1}{2}kF^\sigma_{\;\;\eta}\right)\/\left(-\frac{1}{2}kF^\eta_{\;\;\alpha}e^\alpha_j\right)e^\nu_{p}e^\lambda_i = \\
=
\frac{1}{2}\epsilon^{5plij}\epsilon_{5\nu\rho\lambda\sigma}\/\left(\de{\tilde{\omega}_i^{\lambda\sigma}}/de{x^j} + \tilde{\omega}_{j\;\;\;\eta}^{\;\;\lambda}\tilde{\omega}_i^{\;\;\eta\sigma}\right)\/e^\nu_p  
-\frac{1}{8}\epsilon^{5plij}\epsilon_{5\nu\rho\lambda\sigma}k^2F^{\lambda\sigma}F_{ji}e^\nu_p + \\
-\frac{1}{8}\epsilon^{5plij}\epsilon_{5\nu\rho\lambda\sigma}k^2F^\lambda_{\;\;\alpha}e^\alpha_{j}F^\sigma_{\;\;\beta}e^\beta_{i}e^\nu_p + \frac{1}{8}\epsilon^{5plij}\epsilon_{5\nu\rho\lambda\sigma}k^2F^\sigma_{\;\;\eta}F^\eta_{\;\;\alpha}e^\alpha_{j}e^\nu_{p}e^\lambda_i =0
\end{split}
\end{equation}
By working on the last three terms in the above equations, we get exactly eqs. \eqref{4.11}. Analogous arguments about the equations 
\begin{equation}\label{A.4}
\frac{1}{4}\epsilon^{qplij}\epsilon_{\mu\nu5\lambda\sigma}\/e^\mu_{q}e^\nu_{p}\/\left( \de{\omega_{i}^{\;\;\lambda\sigma}}/de{x^j} + \omega_{j\;\;\;\eta}^{\;\;\lambda}\omega_{i}^{\;\;\eta\sigma} \right) = 0\qquad\quad l\not=5
\end{equation}
(indices from $1$ to $5$). Once again, after some calculations, we have
\begin{equation}\label{A.5}
-\frac{1}{4}\epsilon^{qpli5}\epsilon_{\mu\nu\lambda\sigma5}\/\left(\de{\omega_5^{\;\;\lambda\sigma}}/de{x^i} + \omega_{i\;\;\;\eta}^{\;\;\lambda}\omega_5^{\;\;\eta\sigma}\right)\/e^\mu_{q}e^\nu_{p} + 
\frac{1}{4}\epsilon^{qpli5}\epsilon_{\mu\nu\lambda\sigma5}\omega_{5\;\;\;\eta}^{\;\;\lambda}\omega_i^{\;\;\eta\sigma}e^\mu_{q}e^\nu_{p} =0
\end{equation}
(indices from $1$ to $4$). By inserting the content of \eqref{4.9} and saturating by $e^\alpha_l\/$, we end up with the equations \eqref{4.12}.

\end{document}